# Generation of twin beams from an optical parametric oscillator pumped by a frequency-doubled diode laser

**Kazuhiro Hayasaka[*], Yun Zhang, and Katsuyuki Kasai**

Kansai Advanced Research Center,
National Institute of Information and Communications Technology[**]
588-2 Iwaoka, Iwaoka-cho, Nishi-ku, Kobe, 651-2492 Japan

Quantum-correlated twin beams were generated from a triply resonant optical parametric oscillator with an a-cut KTP crystal pumped by a frequency-doubled diode laser. A total output of 5.1 mW was obtained in the classical-to-nonclassical-light conversion system driven by a 50-mW diode laser at 1080 nm. Quantum noise reduction of 4.3 dB (63%) in the intensity difference between the twin beams was successfully observed at the detection frequency of 3 MHz.

Optical parametric oscillators (OPOs) are popular devices for generating nonclassical light.[1-3] Degenerate operation below threshold can generate various squeezed vacuum states,[1] which are ingredients for recent quantum optics experiments, such as quantum teleportation.[4] Non-degenerate optical parametric oscillators (NOPOs) above threshold can generate quantum-correlated twin beams with quantum correlation in the intensity difference, so-called intensity-difference squeezing.[2,3] One practical advantage of the twin beams is that they are 'bright' in contrast to vacuum squeezed states and, therefore, the detection of the quantum correlation can be relatively easily done by direct detection of the intensity difference.[2,3] This feature has been exploited in applications such as sub-shot-noise measurement,[3] high-sensitivity spectroscopy[5] and secure quantum communication channel.[6] Recent studies report on the nonclassical nature of the twin beams from different aspects.[7-10] Photon-number statistics of twin beams have been revealed by direct detection.[7] Conditional generation of a sub-Poissonian state has been demonstrated by post selection of twin beams.[8] Theoretical studies point out feasible generation of bright entangled beams by self-phase-locked operation of NOPOs.[9,10] These reports make twin beams more important as a resource for quantum information studies. There is a great demand for a practical device to generate twin beams.

A conventional way to generate twin beams is to build a NOPO pumped by a commercial frequency- and intensity-stable, high-power laser. NOPOs consisting of a KTP crystal pumped by a frequency-doubled Nd:YAG laser are often used in this sense.[2,3,5-9] However, the non-zero walk-off angle of KTP for the Type-II phase matching at 1064 nm limits the performance of frequency doubling and parametric oscillation.[11-13] The fundamental wavelength of 1080 nm can solve this problem. It enables Type-II non-critical phase matching (NCPM) in an a-cut KTP crystal.[11-13] This strategy was employed with Nd:YAlO$_3$ (Nd:YAP) lasers by Z. Y. Ou et. al.[11,12] and R. Guo et. al.[13] In the former case frequency doubling with an efficiency as high as 85 % was demonstrated,[11,12] and in the latter case they achieved 5.9 dB (74%) intensity-difference squeezing with an NOPO pumped by an intra-cavity frequency-doubled Nd:YAP/KTP laser.[13] In spite of the excellent performance of KTP at 1080 nm the Nd:YAP laser is not available commercially, and this has been a great barrier for building a practical nonclassical light source. In this letter, we present a more practical setup, which combines the special wavelength and a compact commercial diode laser. The fundamental source is an extended-cavity diode laser (ECDL) at 1080 nm consisting of a single-stripe 50-mW diode laser. It is frequency-doubled in an enhancement cavity with an a-cut KTP crystal to yield 22.8 mW of green light. It is then used to pump a semi-monolithic triply resonant NOPO with an a-cut KTP to generate total output of 5.1 mW. The intensity-difference squeezing of 4.3 dB (63%) was observed. Compared to the other setups,[2,3,5-9,11-13] our setup has two distinct features. First, the setup is the first continuous-wave OPO pumped by a frequency-doubled single-stripe diode laser. Second, it is the first twin-beam generator obtained by direct frequency conversion processes of a single-stripe diode laser. These features have been realized by exploiting NCPM in an a-cut KTP.

---

[*] Corresponding author, e-mail: k.hayasaka@osa.org
[**] Reorganized from Communications Research Laboratory on 1. Apr. 2004

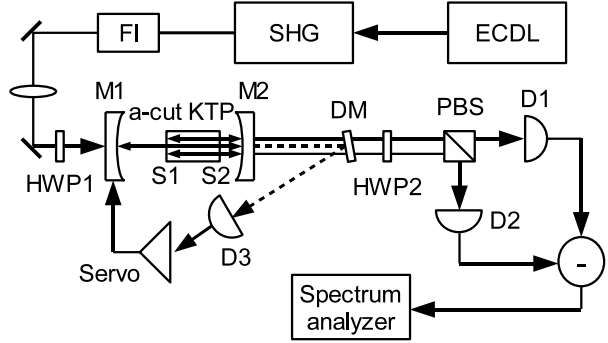

Fig.1 Experimental setup for generation and detection of twin beams. ECDL, extended-cavity diode laser; FI, Faraday isolator; HWP, half-wave plate; DM, dichroic mirror; PBS, polarizing beam splitter.

     Our experimental setup is depicted in Fig.1. Details of the frequency doubler are reported elsewhere.[14] The doubler generates 22.8 mW of green light in a perfect Gaussian mode thanks to the absence of the walk off. The pump beam from the doubler is mode-matched to the OPO by a single lens. An optical isolator with an extinction rate of 35 dB and a half-wave plate are inserted in between the doubler and the OPO. The pump polarization is horizontal at the cavity input port. The harmonic separator guiding the pump beam to the OPO ensures no injection power at 1080 nm leaking out from the doubler to the pump. The OPO consists of two concave mirrors with a curvature of 20 mm, and an a-cut KTP crystal in a dimension of 3 mm x 3 mm x 10 mm. The two mirrors (M1, M2) form a pump cavity, while one of the crystal surface (S1) and one mirror (M2) form a semi-monolithic cavity for the signal and idler modes (signal-idler cavity). For the pump, the transmission of M1 is 3.5%, and M2 is high-reflection (HR) coated. Both surfaces of the crystal are anti-reflection (AR) coated for the pump. For the signal-idler cavity, the crystal surfaces S1 and S2 are HR and AR coated, respectively. A small output transmission of 1.9 % is chosen for M2 to ensure a low threshold power, especially for the initial investigation of this setup. The distance of M1 and M2 are set slightly shorter than the confocal condition, and S1 is located at the beam waist of the pump cavity. The crystal is mounted in a block made of aluminum and its temperature is controlled within ±0.005°C. The mirrors M1, M2 are mounted on ring piezoelectric transducers (PZT) for frequency tuning. The faint leakage of the pump from

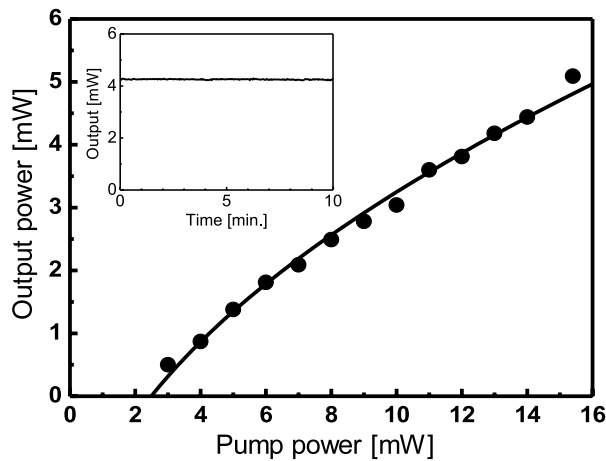

Fig.2 The measured total output power of the OPO as a function of the pump power. The dots show measured output power. The solid curve represents the theoretical fit. The oscillation threshold pump power was determined to be 2.5 mW. The inset shows an example of power stability.

M2 is separated from the output by means of a dichroic mirror (DM). The reflection from this mirror is used to lock the pump cavity to the pump frequency. Since the pump frequency is modulated at 40 kHz in the doubler, this modulation in the pump leakage is used as an error signal by phase-sensitive detection. The error signal is directly fed back to the PZT for the pump cavity mirror M1. Parametric oscillation is obtained by adjusting the signal-idler cavity length by PZT for M2, while the pump cavity is locked to a resonance. In order to obtain the lowest spatial mode, the pump cavity length is first locked to a $TEM_{00}$ pump mode, and then the signal-idler cavity length is adjusted to a $TEM_{00}$ mode of the output. Observation of the far field patterns on a white card and an infrared viewing card was enough to identify the spatial mode. Stable oscillation was established when the voltage applied to the PZT supporting M2 was set to the optimum value. This voltage was found by changing the voltage manually until the output intensity became stable. This adjusting procedure corresponds to finding the cavity configuration satisfying the triple resonance as well as the phase relationship among the pump, signal and idler modes. The signal-idler cavity length was left uncontrolled once the stable oscillation was established. This simple lock scheme kept intensity-stable operation for more than tens of minutes. Typically, the output power stayed within 2% peak-to-peak for more than 10 minutes.

The output transverse mode was diagnosed by the mode pattern on an infrared-viewing card. The both beams operated in the $TEM_{00}$ mode as long as the output power stayed constant. The spectra of the output beams were observed with a confocal Fabry-Perot cavity having a finesse of 80 and a free spectral range of 2 GHz. The signal and the idler beams operated each in a single mode, and the linewidth was 25 MHz limited by the resolution of the Fabry-Perot cavity. The wavelength dependence on the crystal temperature was monitored with an optical spectrum analyzer. For the change from 25 °C to 95 °C, the idler beam changes almost linearly from 1077.8 nm to 1081.5 nm at the rate of ±0.05 nm/°C, while the signal beam changes from 1083.0 nm to 1079.4 nm. The two wavelengths are very close to the degeneracy at 76.2 °C. The following measurement of the power and the noise spectrum was performed at this temperature.

The total output of the OPO was measured with a calibrated power meter. A variable neutral density filter was used to attenuate the pump beam to observe the dependence on the pump power. The measured dependence is shown in Fig.2. Total output power of 5.1 mW was obtained for 16-mW pump power, corresponding to a conversion efficiency of 31.9 %. The whole frequency conversion process from the 44.2-mW ECDL beam to the 5.1-mW twin beams can be interpreted as conversion of the classical light to the nonclassical light with an efficiency of 11.5 %. An example of the intensity stability is shown in the inset to Fig.2. In this data set, the intensity fluctuation was smaller than 1 % peak-to-peak for ten minutes. The threshold pump power was determined by fitting a theoretical curve to the data. The expression $P_{out} = 2\varepsilon ((P_{th} P_p)^{1/2} - P_{th})$ was assumed, where $P_{out}$, $P_p$, $P_{th}$, $\varepsilon$ are output power, pump power, threshold pump power, power slope efficiency, respectively.[15] The fitting gives $P_{th} = 2.5$ mW, $\varepsilon = 0.65$.

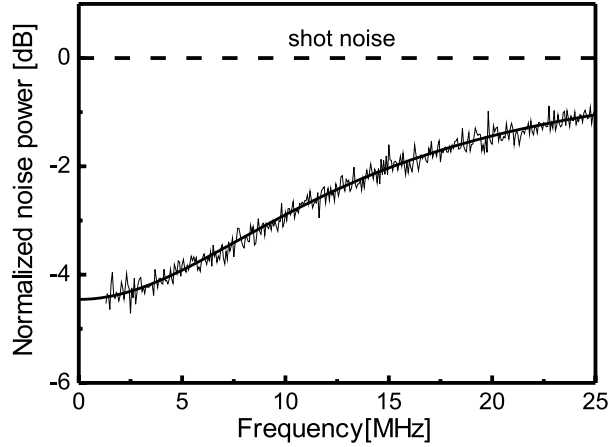

Fig.3 Noise power spectrum of the twin-beam intensity difference. The data are normalized to the shot-noise level indicated by the straight dashed line. The electrical noise was subtracted both from the data and the shot-noise level. The dashed curved line on the data shows the theoretical fit.

Following the characterization of classical operation, the quantum characteristics of the OPO were investigated. The noise spectrum of the twin beams with total power of about 4 mW was measured by the standard method.[2,3,5,13] The beams are first separated by a polarizing beam splitter and then are detected with high-efficiency photo diodes. The power imbalance between the both beams was less than 3 %. The total detection efficiency was estimated to be 90%, in which the detector quantum efficiency is 92%, and the transmission efficiency is 98%.[7] The intensity-difference noise is measured on photocurrent difference, and its spectrum is observed with a radio frequency spectrum analyzer. The measured noise power was taken with sweep time of 2 s, resolution bandwidth of 100 kHz, video bandwidth of 100 Hz. In order to calibrate the shot-noise level corresponding to the total beam power, the half-wave plate (HWP2) is rotated 22.5° respect to the PBS axis and the completely mixed beams were measured.[2] The observed spectrum of the twin-beam intensity difference is shown in Fig. 3. The data are normalized to the shot-noise level. Correction for electrical noise was made both for the data and the shot-noise level before plotting. The intensity-difference squeezing of 4.3 dB (63%) was observed around 3 MHz. At this frequency, the excess noise of the two beams was about 8 dB above the shot-noise level, and electrical noise floor was 5 dB bellow the intensity-difference noise. The dashed line on the data shows the theoretical fit with the Lorenzian curve described by $S(\Omega) = 1 - \eta_d \eta_e / (1 + \Omega^2)$, where $\eta_d$ is the detection efficiency, $\eta_e$ is the escape efficiency, and $\Omega$ is the noise frequency normalized to the cavity bandwidth, respectively.[2,3] The efficiency $\eta_d$ was estimated to be about 90 % in our previous experiment.[7] The escape efficiency $\eta_e$ is expressed by $T / (T + L)$, where $T$ is the transmission of the out-coupling mirror, $L$ is the residual loss in the cavity. The fitting procedure gives $\eta_e = 71$ %, and this suggests $L = 0.78$ %.

The intensity-difference squeezing would be further increased by use of an output coupler with higher transmission. Since the threshold power $P_{th}$ scales as $(T + L)^2$ while the escape efficiency $\eta_d$ is $T / (T + L)$,[2] the possible improvements can be roughly estimated from the intra-cavity loss $L$ estimated above. Although the available pump power is limited to 16 mW with the current mode-matching optics, its optimization would bring pump power of 21 mW to the OPO. This would enable an output coupling of $T = 6$ % for threshold pump power of $P_{th} = $ 16mW, which could lead to the escape efficiency $\eta_e = 88$ % and the intensity-difference squeezing of 6.8 dB (79 %).

In summary we have demonstrated stable continuous-wave operation of a triply resonant OPO pumped by a frequency-doubled diode laser. Intensity-difference squeezing of 4.3 dB at the frequency of 3 MHz was successfully observed in the twin beams generated from the OPO. A conversion efficiency of 11.5 % from classical light to nonclassical light was achieved. Although there are several reports on low-threshold, stable OPOs,[13,16] and also on large intensity-difference squeezing,[2-3,13,17] our setup is distinct from these setups in realizing efficient nonclassical light generation only with easily available optical components. This setup simplifies preparation of nonclassical states of light substantially, and might supply with a chance to experiments demanding multiple modes of nonclassical light.